\documentclass[onecolumn,floatfix,superscriptaddress,showpacs,showkeys,nofootinbib,preprint]{revtex4}%
\usepackage{epsfig}
\usepackage{amssymb,latexsym,amsmath}

\newcommand{\cent}[2]{\mbox{$c$ = #1--#2\%}}

\begin{document}

\title{Transverse-momentum spectra of strange particles produced in Pb+Pb collisions at $\sqrt{s_{\rm NN}}=2.76$~TeV in the chemical non-equilibrium model}

\author{Viktor Begun}
% \email{viktor.begun@gmail.com}
 \affiliation{Institute of Physics, Jan Kochanowski University, PL-25406~Kielce, Poland}
 \affiliation{Bogolyubov Institute for Theoretical Physics, 03680 Kiev, Ukraine}

\author{Wojciech Florkowski}
 %\email{wojciech.florkowski@ifj.edu.pl}
 \affiliation{Institute of Physics, Jan Kochanowski University, PL-25406~Kielce, Poland}
 \affiliation{The H. Niewodnicza\'nski Institute of Nuclear Physics, Polish Academy of Sciences, PL-31342 Krak\'ow, Poland}

\author{Maciej Rybczynski}
 %\email{maciej.rybczynski@ujk.edu.pl}
 \affiliation{Institute of Physics, Jan Kochanowski University, PL-25406~Kielce, Poland}

\date{May 27, 2014}

\begin{abstract}
We analyze the transverse-momentum spectra of strange hadrons produced in Pb+Pb collisions at the collision energy $\sqrt{s_{\rm NN}}=2.76$~TeV. Our approach combines the concept of chemical non-equilibrium with the single-freeze-out scenario. The two ideas are realized in the framework of the Cracow model, whose thermodynamic parameters have been established in earlier studies of the ratios of hadron multiplicities. The geometric parameters of the model are obtained from the fit to the spectra of pions and kaons, only.  Using these parameters, we obtain an excellent description of the spectra of protons and the $K_S^0$, $K^*(892)^0$, and $\phi(1020)$ mesons. A satisfactory description is also obtained for the $\Lambda$, $\Xi$ and $\Omega$ hyperons. Further improvement of the hyperon spectra may be achieved if we assume that they are emitted from a smaller, internal part of the system but at the same thermodynamic conditions. Our work not only includes all particle species measured up to now in heavy-ion collisions at the LHC energies but, in addition, discusses the centrality dependence of the particle production.
\end{abstract}



\maketitle

\section{Introduction}
\label{sect:Intro}

Thermal and statistical models of hadron production
\cite{Koch:1985hk,Cleymans:1992zc,
Cleymans:1998fq,Gazdzicki:1998vd, Braun-Munzinger:1994xr,
Cleymans:1996cd,Becattini:2000jw,
Sollfrank:1993wn,Schnedermann:1993ws,
Braun-Munzinger:1995bp,Becattini:1997uf,
Yen:1998pa,Braun-Munzinger:1999qy,Becattini:2003wp,
Braun-Munzinger:2001ip,Florkowski:2001fp,
Broniowski:2001we,Broniowski:2001uk, Retiere:2003kf} have become the standard tools to analyze mean multiplicities of the particles produced in heavy-ion collisions. They have explained successfully
the AGS \cite{Braun-Munzinger:1994xr,Cleymans:1996cd,
Becattini:2000jw}, SPS \cite{Sollfrank:1993wn,Schnedermann:1993ws,
Braun-Munzinger:1995bp,Becattini:1997uf,
Yen:1998pa,Braun-Munzinger:1999qy,Becattini:2003wp}, and RHIC
\cite{Braun-Munzinger:2001ip,Florkowski:2001fp,
Broniowski:2001we,Broniowski:2001uk,Broniowski:2002wp, Retiere:2003kf} data on hadronic abundances. Supplemented by the proper definition of the spacetime geometry and hydrodynamic flow at freeze-out, the statistical models allow us to describe the transverse-momentum spectra and other soft-hadronic observables \cite{Broniowski:2001we,Broniowski:2001uk,Broniowski:2002wp,
Retiere:2003kf,Florkowski:2010zz}.
Nevertheless, the recent LHC data on heavy-ion collisions show
that the predictions of two popular versions of the statistical
model (the chemical equilibrium model and the strangeness
non-equilibrium model) give too large values for the kaon to pion
ratio, $(K^++K^-)/(\pi^++\pi^-)$, and, especially, for the ratio
of protons to pions $(p+\bar{p})/(\pi^++\pi^-)$
\cite{Abelev:2012wca,Abelev:2013vea}. The recent fit \cite{Stachel:2013zma} gives almost three standard deviations higher values for protons and anti-protons compared to the LHC data.

The pions, kaons and protons are the most abundant particles that are produced in heavy-ion  collisions, therefore, this discrepancy is very uncomfortable for the thermal interpretation of hadron production. The problem with the correct description of protons is sometimes referred to as the {\it proton puzzle} \cite{Rybczynski:2012ed}.

There have been two solutions to this puzzle proposed up to now. The first one is to use the UrQMD model to calculate the modification factors for each particle and then to use these factors in the equilibrium statistical model  to account for possible discrepancies \cite{Steinheimer:2012rd,Becattini:2012xb}. A clear disadvantage of this approach is that adding a hadronic afterburner to a thermal model spoils the natural simplicity of the latter --- the main aim of introducing thermal models was to gain a simple description of hadron production which does not refer directly to the aspects of microscopic dynamics. Nevertheless, using the kinetic model one can successfully attribute the
modification factors to the baryon-antibaryon annihilation and other microscopic mechanisms which are taken into account by the UrQMD simulations.

The second solution to the proton puzzle has been achieved by assuming that statistical hadronization happens out of chemical equilibrium \cite{Petran:2013qla,Petran:2013lja}. An advantage of this approach is that it continues to use simple concepts of thermal approach not invoking directly to kinetic simulations. The mean multiplicities are explained by adding only two non-equilibrium parameters, $\gamma_q$ and $\gamma_s$ (the hadron abundances  scale with $\gamma_q$ and $\gamma_s$ depending on the number of the constituent light and strange quarks within a hadron). It is worth mentioning that the chemical non-equilibrium approach offers also a possibility to calculate the modification factors studied within the UrQMD approach and to interpret them in terms of the non-equilibrium parameters. Such a calculation may form a link between two alternative explanations of the data.

Besides the problems with thermal interpretation of the hadron abundances at the LHC, one encounters also the problems with the hydrodynamic interpretation of the transverse-momentum spectra of pions, kaons and protons, see Fig.~1 in Ref.~\cite{Abelev:2012wca} and Figs. 13, 14, 15 of Ref.~\cite{Abelev:2013vea}. The ratios data/model analyzed in Ref.~\cite{Abelev:2013vea} have a very characteristic convex shape. Quite a few hydrodynamic models are discussed in \cite{Abelev:2013vea}: VISH2+1 \cite{Song:2013qma},
HKM \cite{Karpenko:2012yf}, EPOS~\cite{Werner:2012xh} and Krakow~\cite{Bozek:2012qs}~\footnote{We note that the Krakow hydrodynamic model cited in \cite{Abelev:2013vea} is different from the Cracow
freeze-out model used in our paper, although the former uses also the concept of single-freeze-out at the end of the hydrodynamic evolution.}. All these models exhibit a deficit in the very low-$p_T$ region, with the largest effect of about 25--50\% for pions in the most central collisions (\cent{0}{5}). In the ultra-peripheral collisions (\cent{70}{80}), the models fail to reproduce the data at both the low and high transverse momenta. Similar features appear in the calculations presented in Refs.~\cite{Pang:2013pma} and \cite{Gale:2012rq}. This situation is quite surprising, as the hydrodynamic models are
supposed to work at low momenta, let us say up to $p_T\sim2$~GeV.

In our earlier paper \cite{Begun:2013nga}, we have shown that one can connect the proton puzzle with the anomalous behavior of the pion $p_T$ spectra and solve the two problems within the chemical
non-equilibrium version of the Cracow single freeze-out model. Encouraged by the success of our approach, in this paper we extend our study to all measured hadrons including the recently measured strange mesons and baryons. Following the same procedure as in \cite{Begun:2013nga}, we take the values of {\it thermodynamic} parameters from Ref.~\cite{Petran:2013lja} and find the model {\it geometric} parameters from the fit to the pion and kaon spectra, only. In the next step, with the same parameters we determine the spectra of other hadrons. In this work we show that this strategy leads to an excellent description of the spectra of $p+\bar{p}$, $K_S^0$, $K^*(892)^0$, and $\phi(1020)$. A satisfactory description is also obtained for $\Lambda$, $\Xi$ and $\Omega$. Further improvement of the hyperon spectra may be
achieved if we assume that they are emitted from a smaller, internal part of the system but still at the same chemical non-equilibrium thermodynamic conditions.

In \cite{Begun:2013nga} we analyzed the $p_T$ spectra of pions,
kaons and protons in the most central, \cent{0}{5}, and
semi-peripheral, \cent{30}{40}, collisions. In this paper we
consider all centrality classes from the most central,
\cent{0}{5}, to ultra-peripheral, \cent{80}{90}, collisions and
all particles measured by the ALICE collaboration up to now:
$\pi$, $K$, $p$, $K_S^0$, $\Lambda$, $K^*(892)^0$, $\phi(1020)$,
$\Xi$ and
$\Omega$~\cite{Abelev:2013vea,Abelev:2013xaa,Abelev:2014uua,ABELEV:2013zaa}.

The paper is organized as follows: In Sec.~\ref{sect:TheModel} we introduce our approach based on the Monte-Carlo version of the Cracow model as implemented in the {\tt THERMINATOR} code
\cite{Kisiel:2005hn,Chojnacki:2011hb}. The parameters of the model are fixed by the fits to the pion and kaon spectra, which are described
in Sec.~\ref{sect:fixpar}. The results showing the $p_T$ spectra of strange particles are presented in Sec.~\ref{sect:strange}. The general discussion of our physics results  in Section~\ref{sect:Discussion} and the two Appendices close the paper.

%%%%%%%%%%%%%%%%%%%%%%%%%%%%%%%%%%%%%%%%%%%%%%%%%%%
\section{The Cracow Model}\label{sect:TheModel}
%%%%%%%%%%%%%%%%%%%%%%%%%%%%%%%%%%%%%%%%%%%%%%%%%%%

\subsection{The freeze-out spacetime geometry}

The starting point for our considerations is the Cooper-Frye formula. The hadron rapidity and transverse-momentum distributions are calculated from the expression
\begin{eqnarray}
\frac{dN}{dy d^2p_T} = E \,\frac{dN}{d^3p} = \int d\Sigma \cdot p \, f(p\cdot u),
\label{frye-cooper}
\end{eqnarray}
where $d\Sigma_\mu = \tau_f\, r dr\, d\eta\, d\varphi\, u_\mu$ is an element of the freeze-out hypersurface and $u^\mu$ is the hydrodynamic Hubble-like flow at freeze-out
\begin{eqnarray}
u^\mu ~=~ \frac{x^\mu}{\tau_f}~=~\frac{(t,x,y,z)}{\tau_f}
.\label{Hflow}
\end{eqnarray}
The parameter $\tau_f$ fixes invariant time at freeze-out,
$\tau^2_f = t^2-x^2-y^2-z^2$. Hence, the freeze-out hypersurface
may be conveniently parameterized with the help of three
variables. In our case we use the transverse distance from the
collision axis, $r=\sqrt{x^2+y^2}$, the spacetime rapidity,
$\eta=1/2 \ln(t+z)/(t-z)$, and the  azimuthal angle, $\varphi =
\tan^{-1}(y/x)$. Since we consider a boost-invariant system, the
integration over the spacetime rapidity $\eta$ in
(\ref{frye-cooper}) stretches from minus to plus infinity. On the
other hand, the integration over $r$ is restricted to the range $0
\leq r \leq r_{\rm max}$, where $r_{\rm max}$ defines the edge of
a firecylinder. The  quantities $\tau_f$ and $r_{\rm max}$ are the
only two geometric parameters of the Cracow model.

The distribution function \mbox{$f(p\cdot u)$} consists of primordial (directly produced) and secondary (produced by resonance decays) contributions. The decays are handled by {\tt
THERMINATOR} \cite{Kisiel:2005hn,Chojnacki:2011hb} and include all particles and well established resonances. The primordial distribution of the $i$th hadron in the local rest frame has the form \cite{Torrieri:2004zz}
\begin{eqnarray}
f_i(p,T,\Upsilon_i) ~=~  \frac{g_i}{\Upsilon^{-1}_i
\exp(\sqrt{p^2+m^2_i} / T ) \mp 1}~.
\label{fi}
\end{eqnarray}
Here $g_i$ is the degeneracy factor connected with spin, $m_i$ is the mass of the particle, $\Upsilon_i$ is the particle's fugacity, and $T$ is the system's temperature. The $-1 \, \,(+1)$ sign corresponds to bosons (fermions).

The integration of the distribution function (\ref{fi}) over three-momentum gives the hadron density
\begin{eqnarray}
n_i(T,\Upsilon_i) = \int {d^3p \over (2 \pi)^3} \, f_i(p,T,\Upsilon_i).
\label{share-ni}
\end{eqnarray}
Similarly, the integration of the distribution (\ref{frye-cooper}) over transverse-momentum gives the rapidity distribution $dN/dy$. In the Appendix \ref{app:1} we show that the freeze-out geometry of our model implies the relation~\footnote{We stress that $\eta$ denotes the spacetime rapidity in our model, hence, Eq.~(\ref{dNdydNdeta}) means that the {\it spacetime rapidity} distribution is equal to the {\it rapidity} distribution. On the other hand, the {\it pseudorapidity} and {\it rapidity} densities are usually quite different, especially at $y=0$ \cite{Florkowski:2010zz}.}
\begin{eqnarray}
\frac{dN_i}{dy}=\frac{dN_i}{d\eta} = \pi r_{\rm max}^2\tau_f \, n_i(T,\Upsilon_i).
\label{dNdydNdeta}
\end{eqnarray}
Consequently, the knowledge of the thermodynamic parameters together with the rapidity density allows us to determine the system's volume per unit rapidity
\begin{eqnarray}
\frac{dV}{dy}~=~\pi r_{\rm max}^2\tau_f~.
\label{V}
\end{eqnarray}
As we shall see below, an independent experimental estimate of this quantity may serve us to reduce the number of independent geometric parameters of our model from two to just one.

\subsection{Implementation of the chemical non-equilibrium}

The fugacity factor $\Upsilon_i$ is defined as
\begin{eqnarray}
\Upsilon_i~=~
 \gamma_q^{N^i_q+N^i_{\bar q}} \gamma_s^{N^i_s+N^i_{\bar s}}  \exp \left( \frac{ \mu_Q Q_i + \mu_B B_i  + \mu_S S_i}{T}\right),
 \label{upsiNeq}
\end{eqnarray}
where $\mu_Q$, $\mu_B$, and $\mu_S$ are the electric, baryon, and strange chemical potentials in the system, while $Q_i$, $B_i$, and
$S_i$ are the electric charge, baryon number, and strangeness of the $i$th hadron
\cite{Begun:2013nga,Torrieri:2004zz}. The chemical potentials are very small at the
LHC energies. Therefore, for simplicity~\footnote{This assumption
does not affect our results, because we do not consider very small differences between particles and anti-particles at the LHC and analyze only their total number. We also neglect
the contributions from the charmed hadrons in Eqs.~(\ref{upsiNeq}) and (\ref{upsiNeq1}).}, we set $\mu_Q=\mu_B=\mu_S=0$ and obtain
\begin{eqnarray}
 \Upsilon_i ~=~ \gamma_q^{N^i_q+N^i_{\bar q}}\gamma_s^{N^i_s+N^i_{\bar s}}~.
 \label{upsiNeq1}
\end{eqnarray}
The quantities $N^i_q$ and $N^i_s$ in (\ref{upsiNeq}) and
(\ref{upsiNeq1}) are the numbers of light $(u,d)$ and strange
$(s)$ quarks in the $i$th hadron, while $N^i_{\bar q}$ and
$N^i_{\bar s}$  are the numbers of the antiquarks in the same
hadron. The $\gamma_q$ and $\gamma_s$ parameters account for
deviations from chemical equilibrium and affect the mean
multiplicities of the particles with $(u,d)$ and $(s)$ quarks,
respectively. In this work, similarly to our previous paper
\cite{Begun:2013nga}, we compare two cases: the non-equilibrium
statistical hadronization version of the Cracow model (NEQ SHM),
where $\gamma_q\neq1$ and $\gamma_s\neq1$, and the equilibrium version (EQ SHM), where $\gamma_q=\gamma_s=1$.

In typical calculations, the use of the parameter $\gamma_s$ does not lead to substantial modifications of other thermodynamic parameters, like temperature or volume, but helps to describe strange particles. The appearance of  $\gamma_s$ can be explained, for example, in the so called core-corona model \cite{Becattini:2012sq,Becattini:2008yn,Becattini:2008ya,Bozek:2005eu},
where a superposition of two sources of particle production is taken into account: single nucleon-nucleon (NN) collisions and a fully equilibrated source.

The use of $\gamma_q>1$ (to our knowledge, introduced for the first time in \cite{Letessier:1998sz}) makes the freeze-out temperature and/or volume smaller, because it
influences the most abundant particles in the medium: pions are multiplied by $\gamma_q^2$ while protons by $\gamma_q^3$. The
temperature found in the recent chemical non-equilibrium
calculations \cite{Petran:2013lja} is about $140$~MeV. It is lower
than the transition temperature obtained by the Wuppertal-Budapest
Collaboration, $T_c=$ 150--170 MeV. We note, however, that direct
comparisons of the chemical non-equilibrium models (i.e., the
models with $\gamma_s \neq 1$ and/or $\gamma_q \neq 1$) with the
lattice simulations is inappropriate, since the lattice
simulations are done for full chemical equilibrium. Furthermore,
the temperature of freeze-out may be not connected with the phase
transition temperature --- we expect only that the latter is
higher than the former.

It is also worth emphasizing that the case with
$\gamma_q\neq1$ and $\gamma_s\neq1$ is equivalent to the introduction of
the non-equilibrium chemical potentials  $\mu_q/T=\ln\gamma_q$ and
$\mu_s/T=\ln\gamma_s$ through the relation
\begin{eqnarray}\label{upsiNeq2}
 \Upsilon_i ~\equiv~ \exp\left(\frac{\mu_q\left(N_q^i+N_{\bar{q}}^i\right)+\mu_s\left(N_s^i+N_{\bar{s}}^i\right)}{T}\right).
\end{eqnarray}
From Eq.~(\ref{upsiNeq2}) one can conclude, for example, that the
conditions $\mu_i>0$ or $\gamma_i>1$ \mbox{($i=q,s$)} mean that
the number of quark and anti-quark pairs in this case is larger
than the corresponding equilibrium number obtained with the same
temperature. This kind of phenomenon may appear because of fast
expansion and cooling of the strongly interacting system. It can
be also a result of the interplay between annihilation and
recombination processes and, possibly, QCD mechanisms like the
gluon condensation followed by the formation of low momentum
$q\bar{q}$ pairs which fuse into pions which subsequently condense
\cite{Letessier:1998sz}, see also
\cite{Blaizot:2011xf,Blaizot:2013lga}. Nevertheless, using
Eqs.~(\ref{upsiNeq1}) and (\ref{upsiNeq2}) we imply that there are
only two parameters responsible for deviations from the standard
statistical model --- the corrections for all particles scale in
the way corresponding to their quark content.

We note that the value of $\gamma_q$ used in
Ref.~\cite{Petran:2013lja} is equivalent to the pion chemical potential $\mu_{\pi}=2T\ln\gamma_q\simeq 134~$MeV, which is very close to the $\pi^0$ mass, $m_{\pi^0}\simeq134.98$. It may suggest that a substantial part of $\pi^0$ mesons form the condensate. Since the prediction of the Bose condensation in 1924~\cite{Bose:1924mk,Einstein} the beauty and simplicity of this phenomenon have attracted attention of many
physicists~\cite{Zimanyi:1979ga,Mishustin:1992nk,Greiner:1993jn,Pratt:1993uy,Florkowski:1995hv,Csorgo:1997us,Bialas:1998jr,Lednicky:1999xz,Begun:2006gj,Blaizot:2011xf,Kokoulina:2013eta,Blaizot:2013lga}.
However, only recently it has been confirmed experimentally in the system of cold atoms~\cite{Anderson:1995gf,Davis:1995pg}. The high
temperature Bose condensation on the MeV$\sim 10^{12}$K scale is also possible even in very small systems which are created in elementary particle collisions~\cite{Begun:2008hq}. The Bose condensate formed in the ultra-relativistic regime has been
considered in Refs.~\cite{Madsen:1992am,Madsen:2000zd,Boyanovsky:2007ay} as a dark matter candidate in cosmological models. There are also interesting effects that appear inside of the pion condensate, see, 
e.g., Ref.~\cite{Kalaydzhyan:2014bfa}. Besides that, large pion chemical potentials may lead to the formation of other types of condensates like a di-quark Bose condensate~\cite{Rapp:1997zu,Splittorff:2000mm}. All those findings indicate at the importance of further studies of the Bose condensation phenomenon in high-energy physics.

%%%%%%%%%%%%%%%%%%%%%%%%%%%%
\begin{figure}[t]
\begin{center}
\includegraphics[width=0.45\textwidth]
{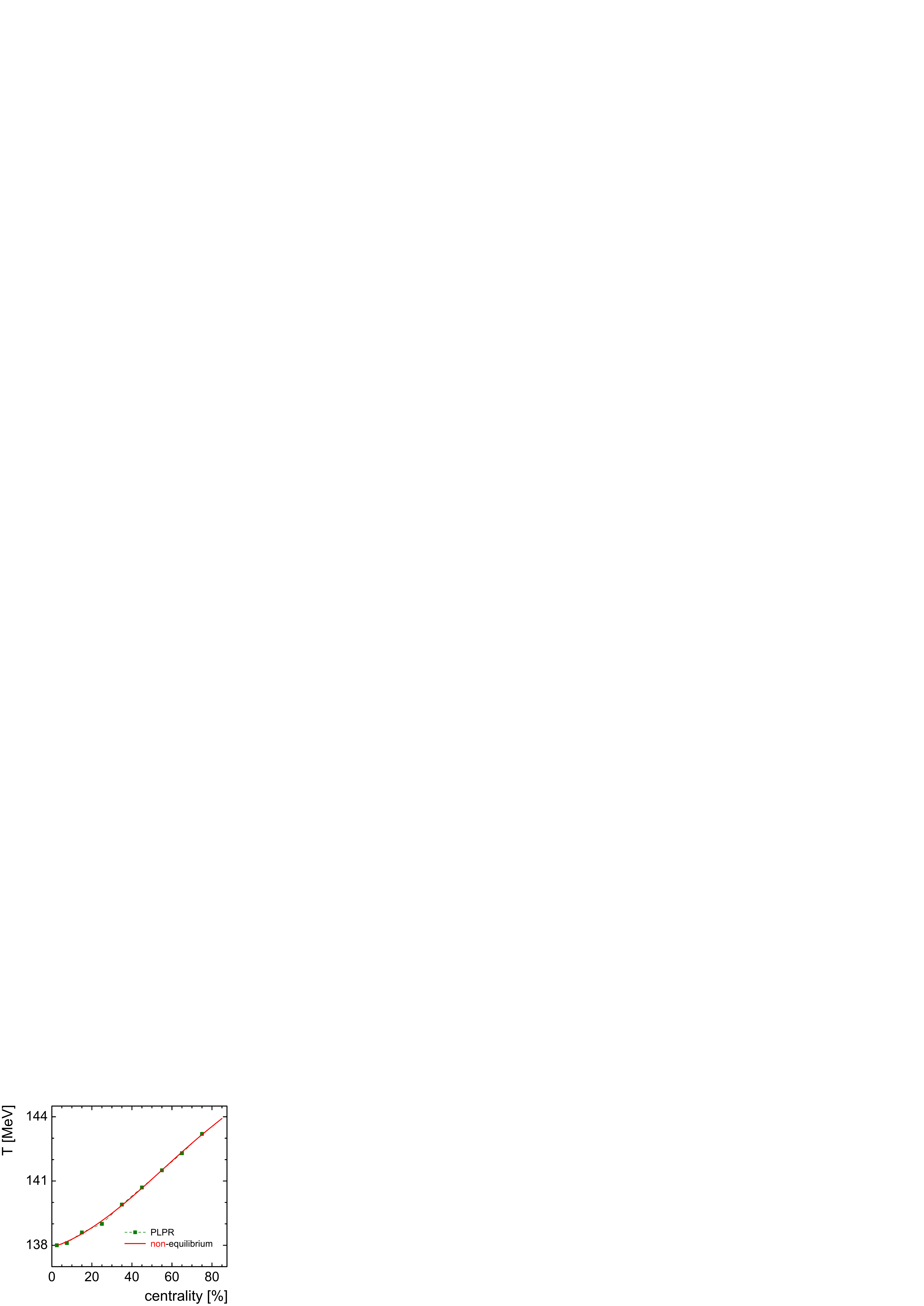}
\end{center}
\caption{(Color online) Temperature $T$ shown as a function of centrality for the chemical non-equilibrium model (solid red line). The results from  Ref.~\cite{Petran:2013lja} are denoted as PLPR.} \label{fig:T}
\end{figure}

%%%%%%%%%%%%%%%%%%%%%%%%%%%%%%%%%%%%%%%%%%%%%%%%
\section{Fixing model parameters --- spectra of pions and kaons}
\label{sect:fixpar}
%%%%%%%%%%%%%%%%%%%%%%%%%%%%%%%%%%%%%%%%%%%%

Similarly as in our previous work \cite{Begun:2013nga}, we use the thermodynamic parameters of the NEQ SHM model determined first in Ref.~\cite{Petran:2013lja}. In Ref.~\cite{Begun:2013nga} we used the values of $T$, $\gamma_q$, and $\gamma_s$ from \cite{Petran:2013lja} and determined the values of $r_{\rm max}$ and $\tau_f$ from the $\chi^2$ fit to the spectra of pions and kaons.  In this work we adapt a simpler method --- in addition to $T$, $\gamma_q$, and $\gamma_s$ we use also the value of the volume $dV/dy$ determined in \cite{Petran:2013lja}. The latter introduces a relation between $r_{\max}$ and $\tau_f$, see Eq.~(\ref{V}), hence, we need to fit only one parameter, which we choose to be the ratio $r_{\rm max}/\tau_f$. In multiple calculations we have verified that the use of $T$, $\gamma_q$, and $\gamma_s$ from \cite{Petran:2013lja} together with the two-dimensional fit of $r_{\rm max}$ and $\tau_f$ leads to the same results as the use of $T$, $\gamma_q$, $\gamma_s$, and $dV/dy$ from \cite{Petran:2013lja} along with the one dimensional fit of $r_{\rm max}/\tau_f$.

In order to analyze the centrality classes different from those studied in Ref.~\cite{Petran:2013lja} and to facilitate the numerical manipulations, we use polynomial approximations for the functions $T(c)$, $\gamma_q(c)$, $\gamma_s(c)$, and $dV/dy(c)$. They are explicitly given in Appendix \ref{app:2}. In practice, for the centrality class defined by the range \cent{$c_1$}{$c_2$}, we use the values from the middle of the range, for example, we take $T((c_1+c_2)/2)$. Having determined the optimal value of $r_{\rm max}/\tau_f$ for each studied centrality class,  we use it to make predictions for the $p_T$ spectra of protons and other hadron species.

%%%%%%%%%%%%%%%%%%%%%%%%%%%%
\begin{figure}[t]
\begin{center}
\includegraphics[width=0.45\textwidth]
{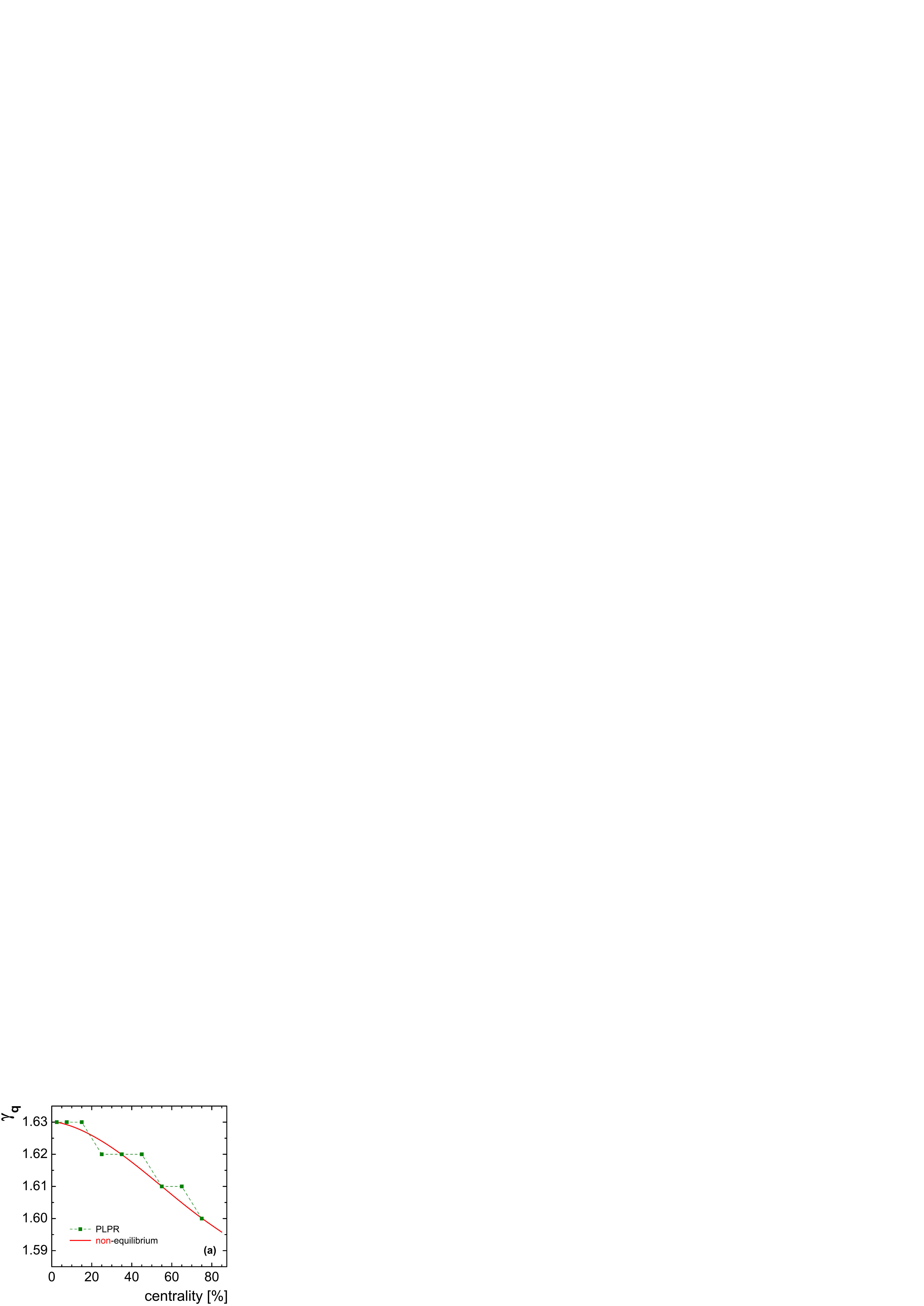}
\includegraphics[width=0.45\textwidth]
{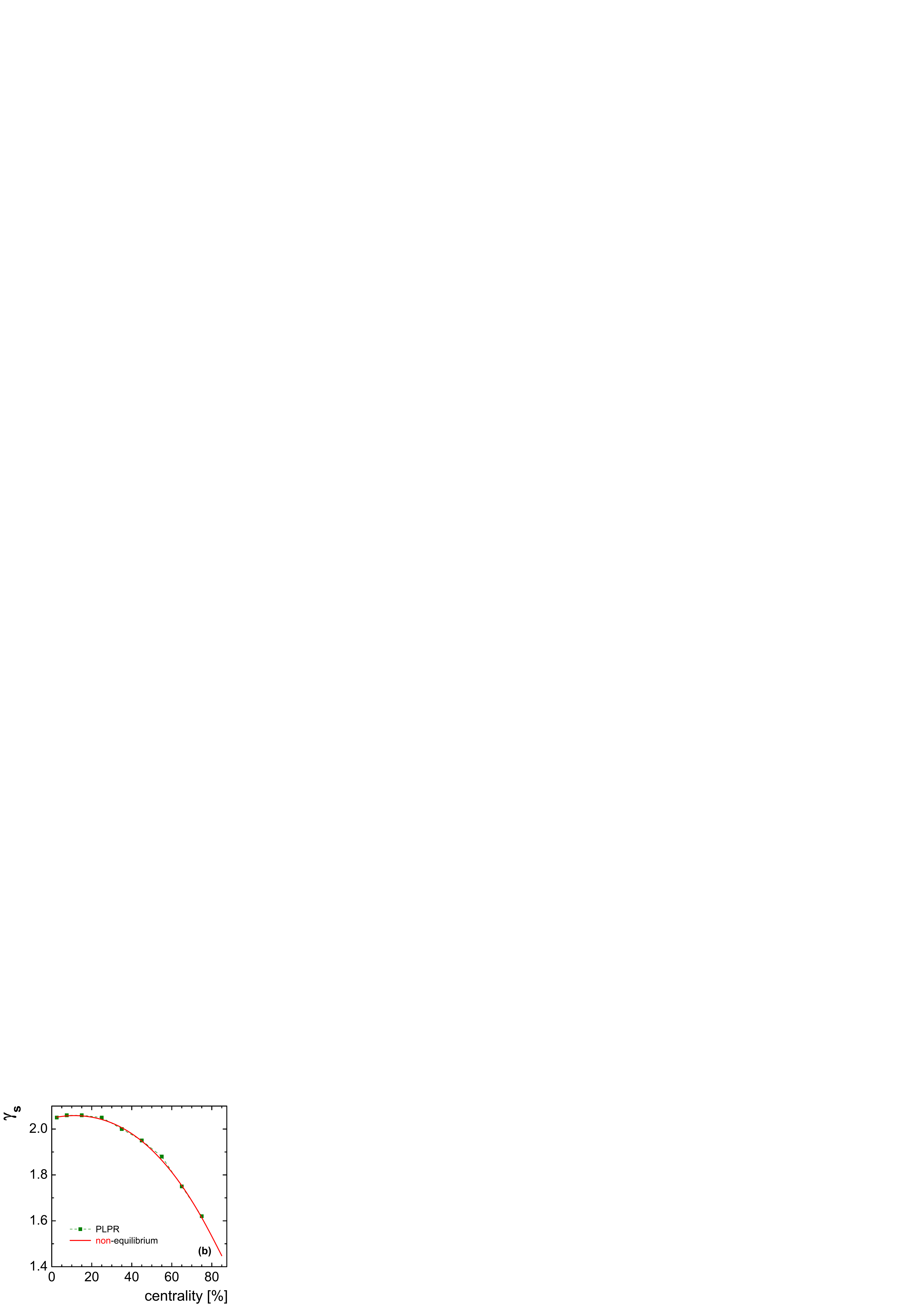}
\end{center}
\caption{(Color online) The parameters $\gamma_q$ (a) and $\gamma_s$ (b) as functions of centrality for the chemical non-equilibrium model.} \label{fig:gammas}
\end{figure}

%%%%%%%%%%%%%%%%%%%%%%%%%%%%
\begin{figure}[t]
\begin{center}
\includegraphics[width=0.45\textwidth]
{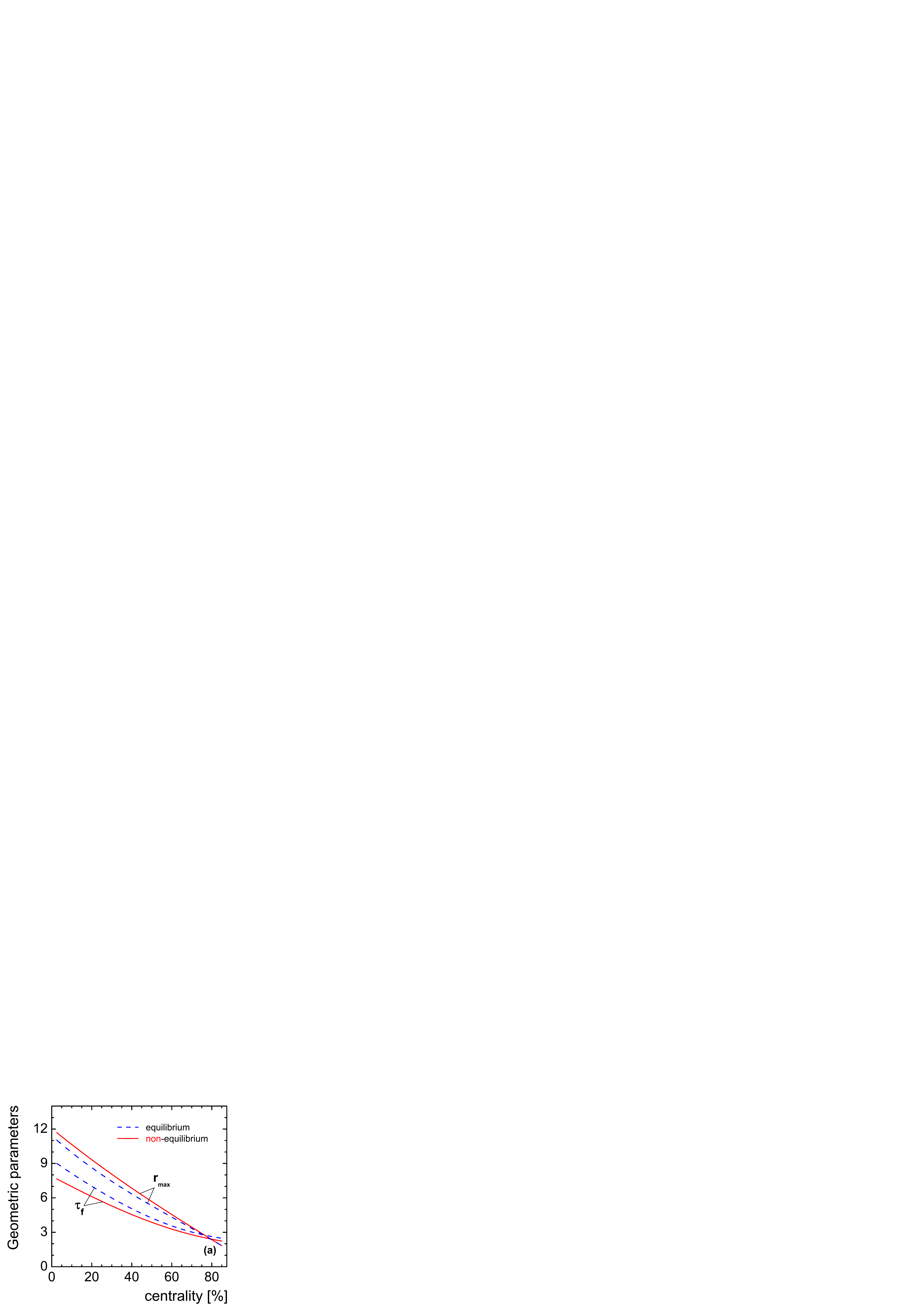}
\includegraphics[width=0.45\textwidth]
{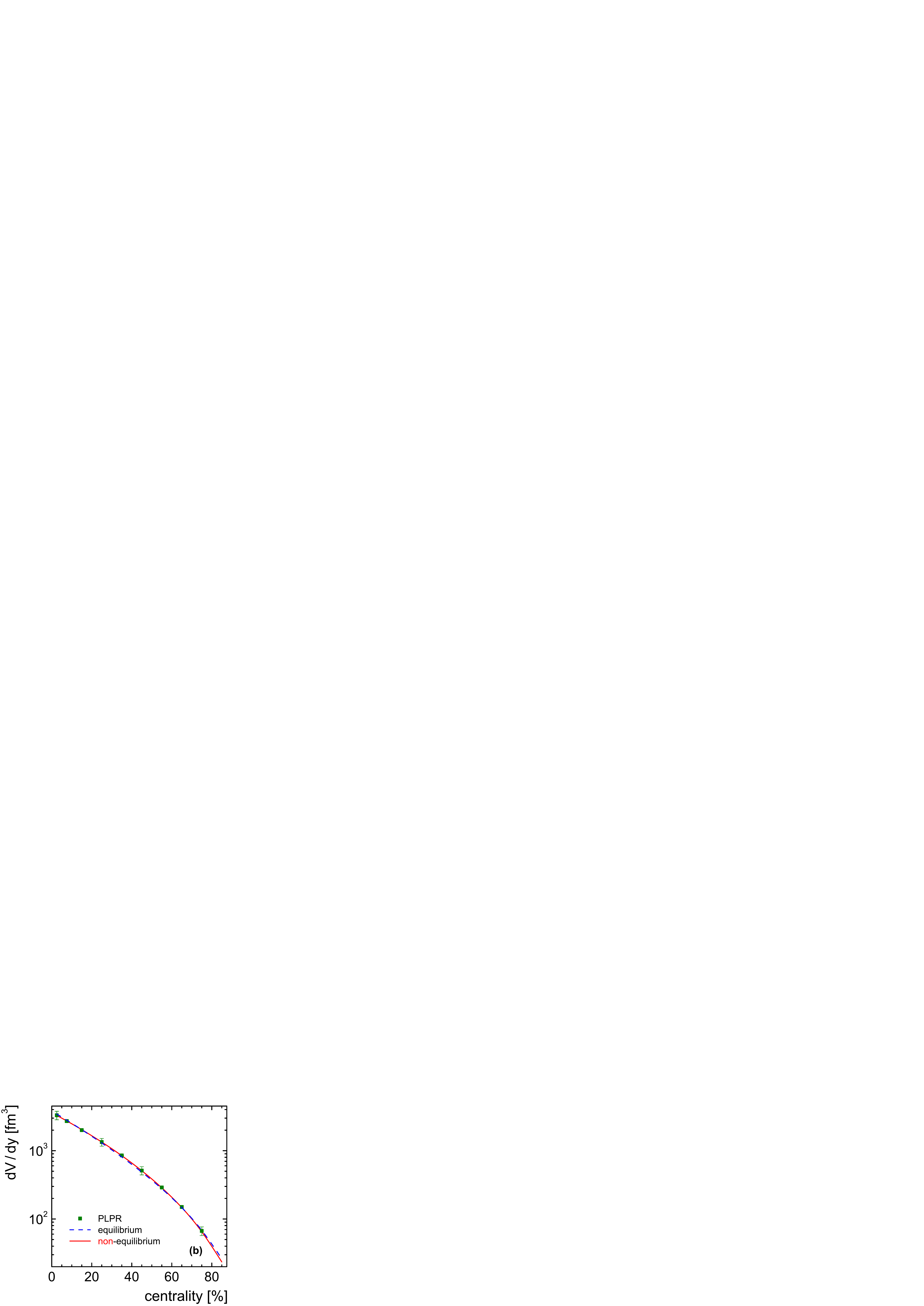}
\end{center}
\caption{(Color online) Geometric parameters (a) and the volume per unit rapidity (b) for both the chemical non-equilibrium and chemical equilibrium models.} \label{fig:GV}
\end{figure}

In Fig.~\ref{fig:T} we show the centrality dependence of the temperature $T$ used in NEQ SHM. The small squares represent the values taken from Ref.~\cite{Petran:2013lja}, the dashed line (denoted as PLPR) is the interpolation of the results found in Ref.~\cite{Petran:2013lja}, and the red line represents our approximation. In Fig.~\ref{fig:gammas} we show the two analogous plots of $\gamma_q(c)$ and $\gamma_s(c)$ in the (a) and (b) panels, respectively. In Fig.~\ref{fig:GV} we show the geometric parameters (a) and the volume per unit rapidity (b) for both the chemical non-equilibrium and chemical equilibrium models. In the EQ SHM version we fix the temperature to be the same for all centralities, $T=165.6$~MeV, and fit both $r_{\rm max}$ and $\tau_f$. In this case, once again we fit first the pion and kaon spectra only, and use the obtained parameters for all other particles. It is interesting to notice that although the geometric parameters are different for NEQ SHM and EQ SHM the volume per unit rapidity remains almost unchanged if we fix the centrality class.

%%%%%%%%%%%%%%%%%%%%%%%%%%%%
\begin{figure}[t]
\begin{center}
\includegraphics[angle=0,width=0.95\textwidth]
{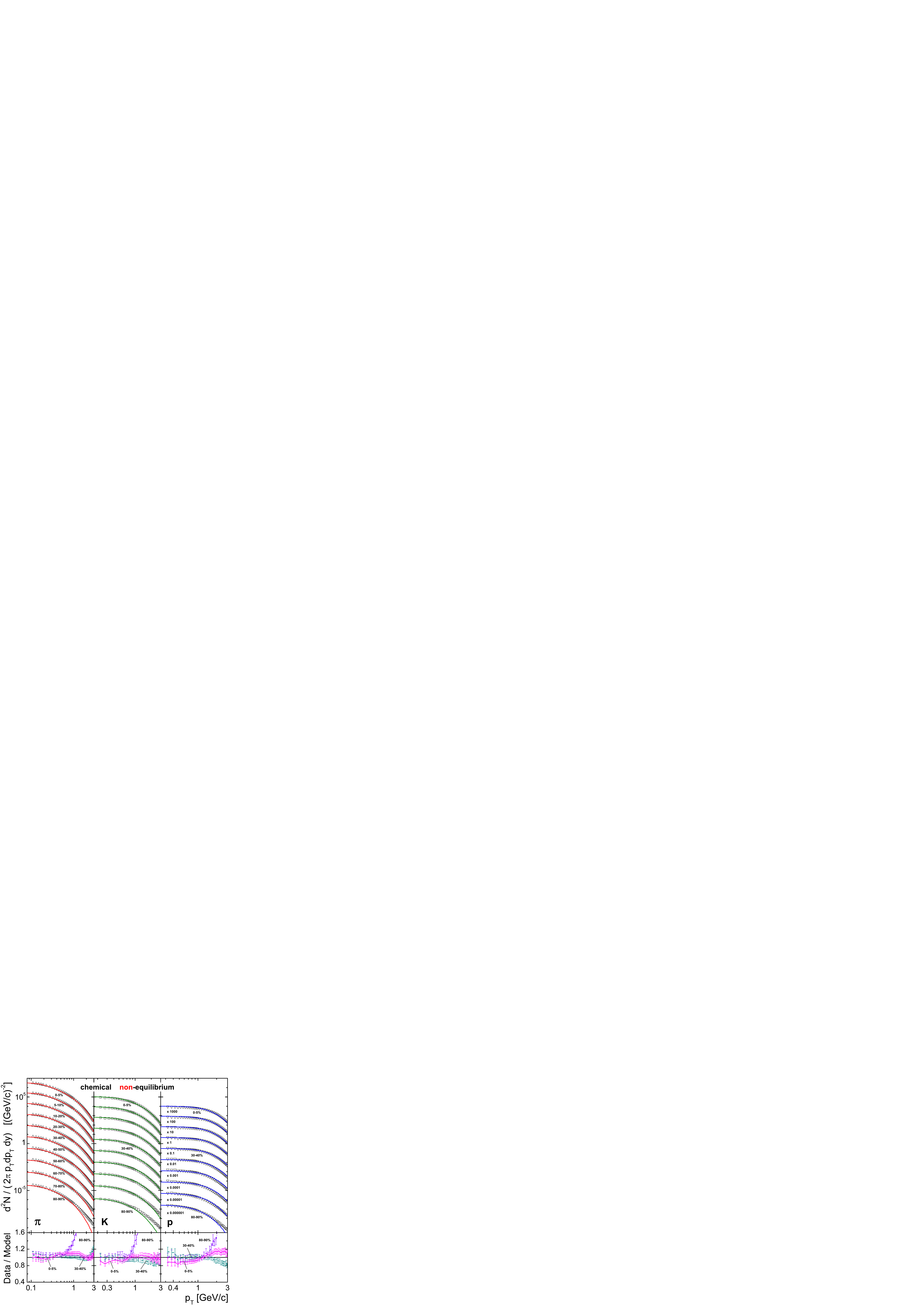}
\end{center}
\caption{(Color online)
 Upper panels: Transverse-momentum spectra
of pions (left), kaons (middle) and protons (right) in different centrality classes. The data \cite{Abelev:2013vea} are shown by
the open symbols. The calculations in the non-equilibrium version of the Cracow model are indicated by the lines. Lower panels: The ratios of the experimental and theoretical $p_T$ spectra in the most central (\cent{0}{5}), semi-peripheral (\cent{30}{40}) and ultra-peripheral (\cent{80}{90}) collisions for pions, kaons, and
protons.} \label{fig:fig1}
\end{figure}
%%%%%%%%%%%%%%%%%%%%%%%%%%%%

%%%%%%%%%%%%%%%%%%%%%%%%%%%%
\begin{figure}[t]
\begin{center}
\includegraphics[angle=0,width=0.95\textwidth]
{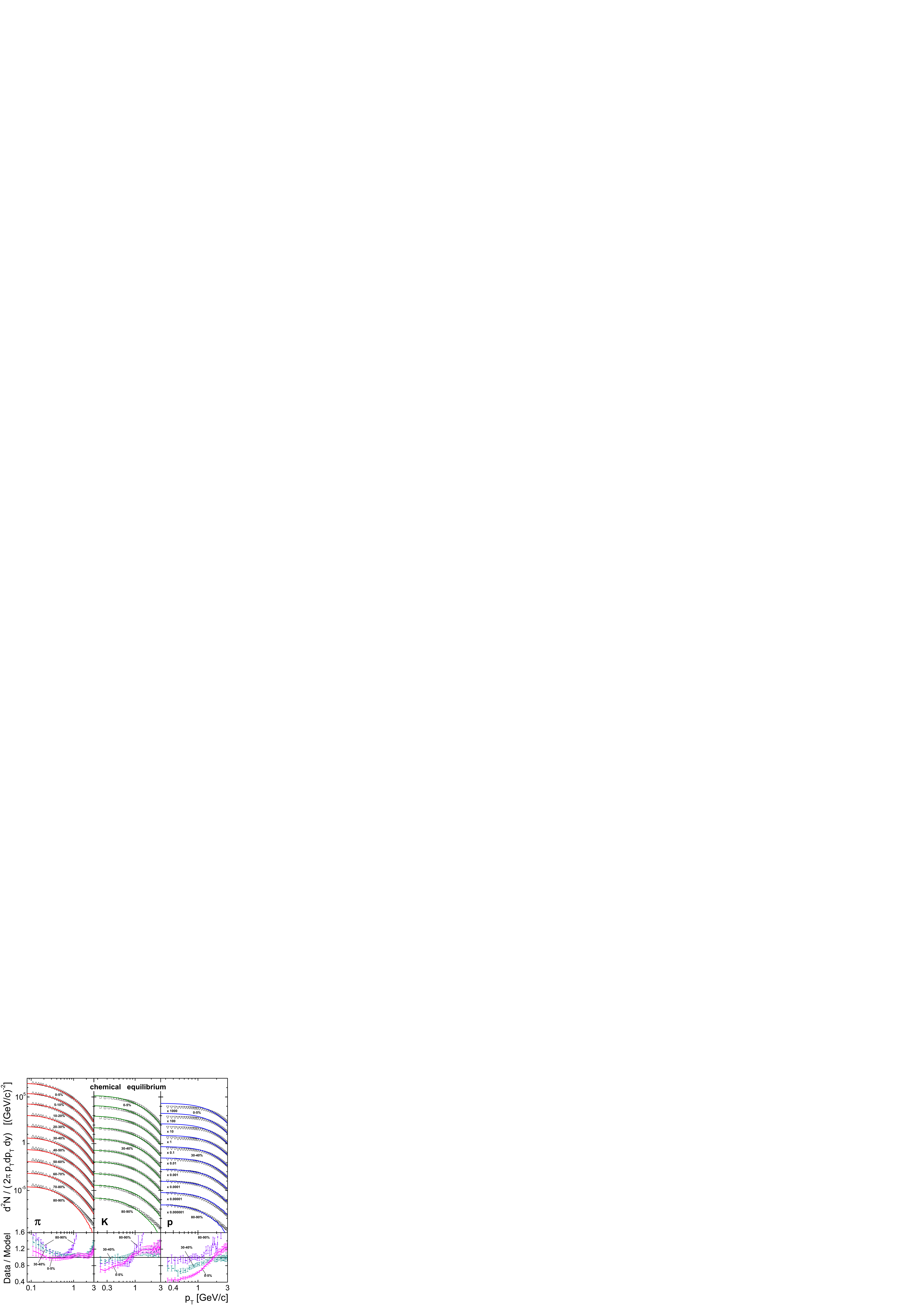}
\end{center}
\caption{(Color online) Same as Fig.~\ref{fig:fig1} but in the equilibrium Cracow model.} \label{fig:fig2}
\end{figure}
%%%%%%%%%%%%%%%%%%%%%%%%%%%%

The results of our calculations for pions, $\pi^++\pi^-$, kaons, $K^++K^-$, and protons,  $p+\bar{p}$, are shown in Fig.~\ref{fig:fig1} for NEQ SHM and in Fig.~\ref{fig:fig2} for EQ SHM. We consider all centralities and the whole $p_T$ range provided by the experiment for pions and kaons. In order to show all centralities together we have multiplied each spectrum at a given centrality by the factor displayed in the upper right panel. Experimental error bars in the upper panels are of the size of the symbols and, therefore, they are not shown. The logarithmic scale used for the $p_T$-axis emphasizes the low $p_T$ region. The lower panels show the data to model ratios for the most central, \cent{0}{5}, semi-peripheral, \cent{30}{40}, and ultra-peripheral collisions, \cent{80}{90}.

In the upper panels of Figs.~\ref{fig:fig1} and \ref{fig:fig2} one can observe a good agreement for pions and kaons both for NEQ and EQ versions of the Cracow model in the wide $p_T$ range and for all centralities. This agreement is a strong argument in favor of the parametrization (\ref{Hflow}) of the flow at freeze-out. On the other hand, the protons in central collisions are described
only in NEQ, as we first observed in \cite{Begun:2013nga}.

The agreement between the data and the model predictions is more clearly displayed in the lower panels of Figs.~\ref{fig:fig1} and \ref{fig:fig2} where the linear vertical scale is used. The NEQ
lines in the Cracow model go exactly through the experimental points for pions and kaons, for most central and semi-peripheral collisions in the whole range from the lowest available point up to $p_T=3$~GeV. The deviations appear only in ultra-peripheral collisions for $p_T \gtrsim 1.5$~GeV. In spite of the fact that we fitted only pions and kaons, the agreement for protons is also very good.

Comparing Figs.~\ref{fig:fig1} and \ref{fig:fig2} one can check that the NEQ fit is much better than the EQ fit. Moreover, in EQ SHM the demand of the best fit for pions and kaons bends pions up and kaons down at low $p_T$. The proton spectra behave similarly to the kaon spectra. The protons are so much in anti-correlation with the pions, that a simultaneous fit of the low $p_T$ part of the spectrum of pions and protons in EQ SHM seems to be impossible.

%%%%%%%%%%%%%%%%%%%%%%%%%%%%
\begin{figure}[t]
\begin{center}
\includegraphics[angle=0,width=0.5\textwidth]
{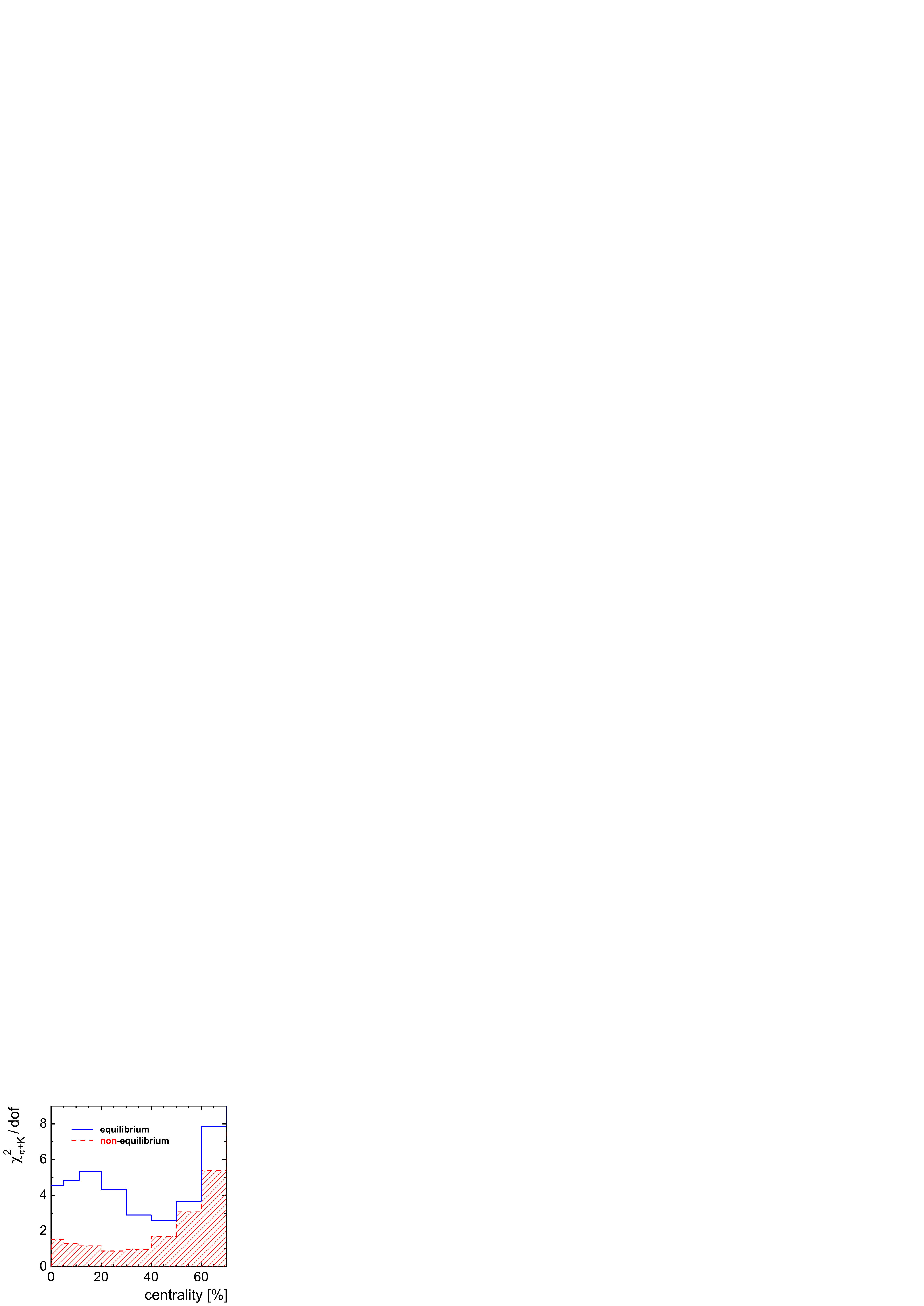}
\end{center}
\caption{(Color online) The centrality dependance of the $\chi^2$
calculated for the joint fit of the $p_T$ spectra of pions and
kaons.} \label{fig:Chi2}
\end{figure}
%%%%%%%%%%%%%%%%%%%%%%%%%%%%

The quality of the fit in NEQ and EQ is illustrated in Fig.~\ref{fig:Chi2}. The values of $\chi^2$ indicate that NEQ SHM is three times better for central and semi-central collisions. Starting with the centrality of about $40\%$,  the difference between NEQ and EQ SHM  decreases while the values of $\chi^2$ grow very rapidly. This behavior may be explained by the qualitative change of the spectra which are exponential in central and semi-central collisions and become well described by a power law in peripheral collisions.

%%%%%%%%%%%%%%%%%%%%%%%%%%%%%%%%%%%%%%%%%%%%%%%%
\section{Spectra of strange particles}
\label{sect:strange}
%%%%%%%%%%%%%%%%%%%%%%%%%%%%%%%%%%%%%%%%%%%%

Another challenging test for the NEQ SHM model is a comparison of the model predictions with the data available for the $p_T$ spectra of strange particles. In order to verify the model we use the same parameters as those found in the study of pions and kaons. We do not present here the results for the chemical equilibrium version, since it always yields much worse agreement with the data as compared to the chemical non-equilibrium version.

In the model analysis of the spectra of strange particles it is very important to take into account the same weak decay corrections as those considered by the experiment. The ALICE Collaboration does not specify the weak corrections for $K^0_S$, hence, we take into account the $K^0_S$'s coming from all possible decays. The spectra of the $\Lambda$ hyperons were corrected for the feed-down  contributions coming from the weak decays of $\Xi^{-}$ and $\Xi^{0}$. Therefore, we subtract the feed-down from these particles only. ALICE also did not correct the $\Lambda$ spectra for the feed-down from non-weak decays of $\Sigma^0$ and from the $\Sigma(1385)$ family. All $\Sigma^0$ and $88.25\%$ of $\Sigma^-(1385)$, $\Sigma^0(1385)$ and $\Sigma^+(1385)$ decay into $\Lambda$, therefore, we include all of them in the $\Lambda$ yield. The $\Xi$'s and $\Omega$'s are directly taken from the generated events.

%%%%%%%%%%%%%%%%%%%%%%%%%%%%
\begin{figure}[t]
\begin{center}
\includegraphics[angle=0,width=0.95\textwidth]
{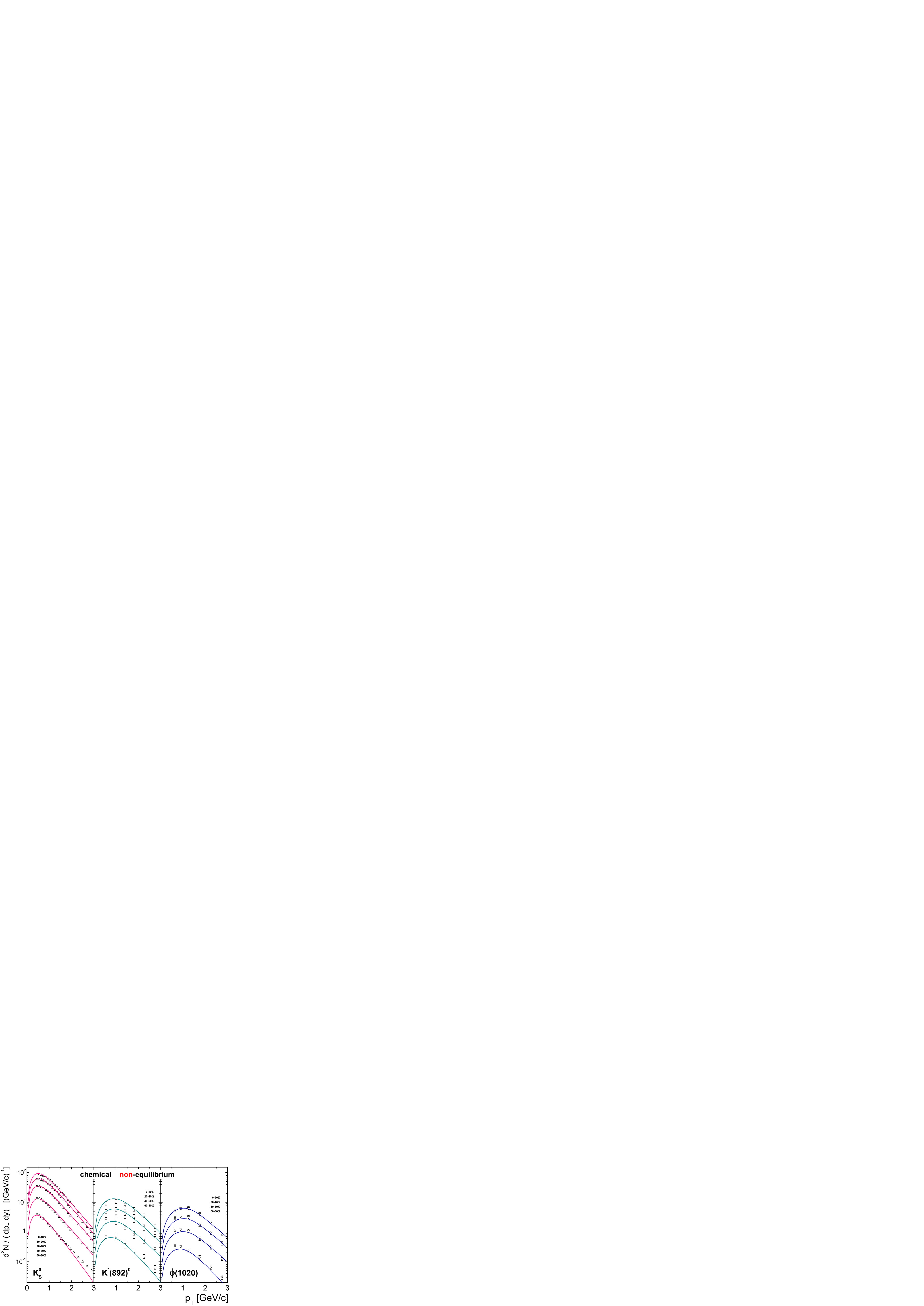}
\end{center}
\caption{(Color online) Transverse-momentum spectra of $K^0_S$
(left), $K^*(892)^0$ (middle) and $\phi(1020)$ (right) in
different centrality classes. The data
\cite{Abelev:2013xaa,Abelev:2014uua} are shown by the open symbols. The calculations in the non-equilibrium Cracow model are indicated by the lines.} \label{fig:K0S_Kstar_Phi}
\end{figure}
%%%%%%%%%%%%%%%%%%%%%%%%%%%%

The results for $K^0_S$, $K^*(892)^0$ and $\phi(1020)$ are shown in Fig.~\ref{fig:K0S_Kstar_Phi}. The error bars are indicated
only if they are bigger than the corresponding symbols in the
figure. One can see that the $p_T$ spectra of these particles are
fitted very well. For many centralities the NEQ model lines go through the experimental points. We stress that it is very
nontrivial that the fit done initially for $\pi^++\pi^-$ and
$K^++K^-$ only appears so good also for $p+\bar{p}$, $K^0_S$,
$K^*(892)^0$ and $\phi(1020)$. These particles have a different
quark content and therefore different non-equilibrium corrections
according to Eq.~(\ref{upsiNeq1}). Moreover, the $K^*(892)^0$, in
contrast to other particles, is a short living resonance that
could interact frequently with the hadronic matter possibly formed
in the final state. The fact that we fit its spectrum together
with the long living $\phi(1020)$ supports our picture of  the
non-equilibrium hadronization and the single freeze-out.

%%%%%%%%%%%%%%%%%%%%%%%%%%%%
\begin{figure}[t]
\begin{center}
\includegraphics[angle=0,width=0.95\textwidth]
{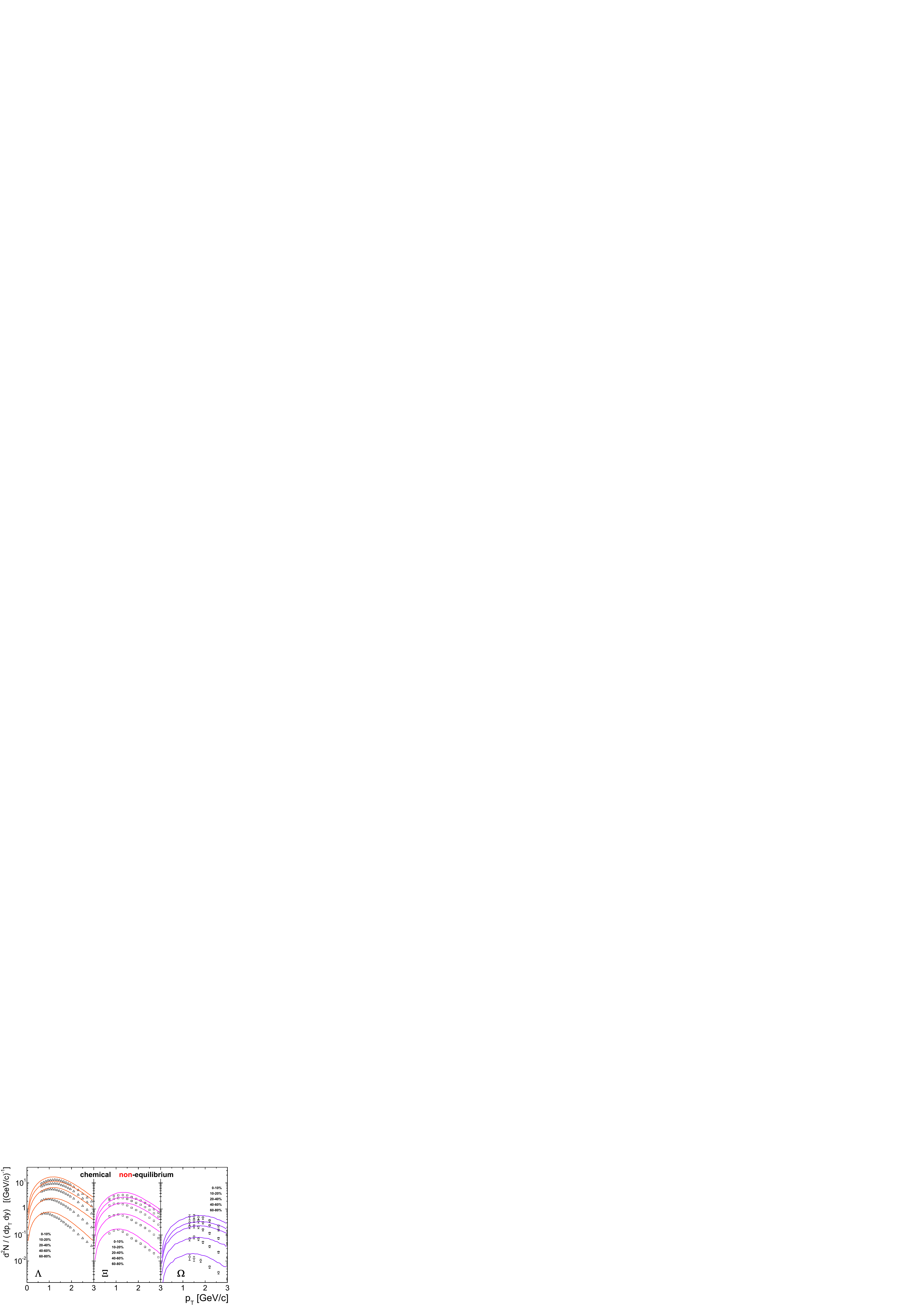}
\end{center}
\caption{(Color online) The same as in
Fig.~\ref{fig:K0S_Kstar_Phi} but for $\Lambda$ \cite{Abelev:2013xaa}, $\Xi$ and $\Omega$ \cite{ABELEV:2013zaa}.} \label{fig:LXiOmAll}
\end{figure}
%%%%%%%%%%%%%%%%%%%%%%%%%%%%

The model and experimental $p_T$ spectra of the hyperons
($\Lambda$, $\Xi=\Xi^++\Xi^-$, and $\Omega=\Omega^++\Omega^-$) are
shown in Fig.~\ref{fig:LXiOmAll}. One can see that the
experimental results at low $p_T$ are reproduced, but for higher
values of $p_T$ ($p_T >$ 2 GeV) the NEQ SHM overshoots the data.
Such deviations increase for heavier particles. On one hand, we
expect that our model's predictions may break down at large values
of the transverse momentum, on the other hand, the observed
differences can be an artefact of the Cracow model that assumes a
simple Hubble form of flow at freeze-out for all particles, see
Eq.~(\ref{Hflow}). Thus, the excess at high $p_T$ may be an
indication that heavy particles in our model  experience too much flow~\footnote{A similar conclusion was drawn from the blast wave
fits presented in Ref.~\cite{Melo+Tomasik}}.

\begin{figure}[t]
\begin{center}
\includegraphics[angle=0,width=0.95\textwidth]
{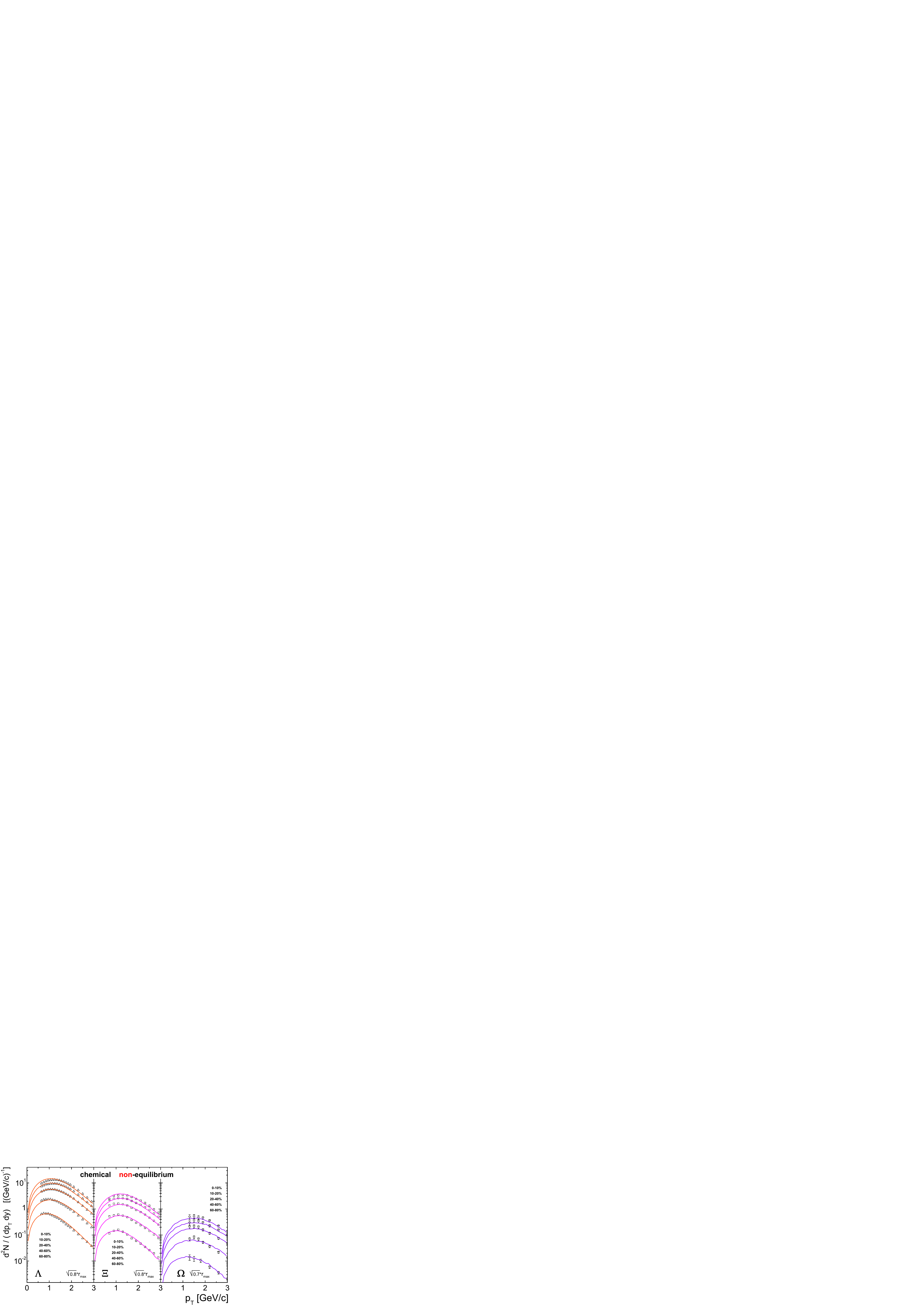}
\end{center}
\caption{(Color online) The same as in  Fig.~\ref{fig:LXiOmAll} but assuming the emission of $\Lambda$, $\Xi$ and $\Omega$ from the
inner part of the system, see text for details.}
\label{fig:LXiOmAlldV}
\end{figure}

We have checked, that one can improve the agreement between the
model predictions and the data by using the same freeze-out time
and choosing a different maximum radius of the firecylinder for
heavy strange particles. It changes their multiplicity which is
proportional to the volume $dV/dy \sim r_{\rm max}^2$ and the
shape of the spectrum, that is sensitive to the ratio $r_{\rm
max}/\tau_f$. The assumption of a smaller emission volume for
$\Lambda$'s and $\Xi$'s (by 20\%) and also for $\Omega$'s (by
30\%) gives us a remarkable agreement, see
Fig.~\ref{fig:LXiOmAlldV}. With the reduced $r_{\rm max}$ and other parameters unchanged, the model results agree with the experimental points even for $\Omega$'s measured at the highest measured centrality. To some extent, such an approach reminds the recent proposal of a two step freeze-out~\cite{Chatterjee:2013yga}.
%
%%%%%%%%%%%%%%%%%%%%%%%%%%%%
%%%%%%%%%%%%%%%%%%%%%%%%%%%%
%

%%%%%%%%%%%%%%%%%%%%%%%%%%%%
\begin{figure}[t]
\begin{center}
\includegraphics[width=0.7\textwidth]
{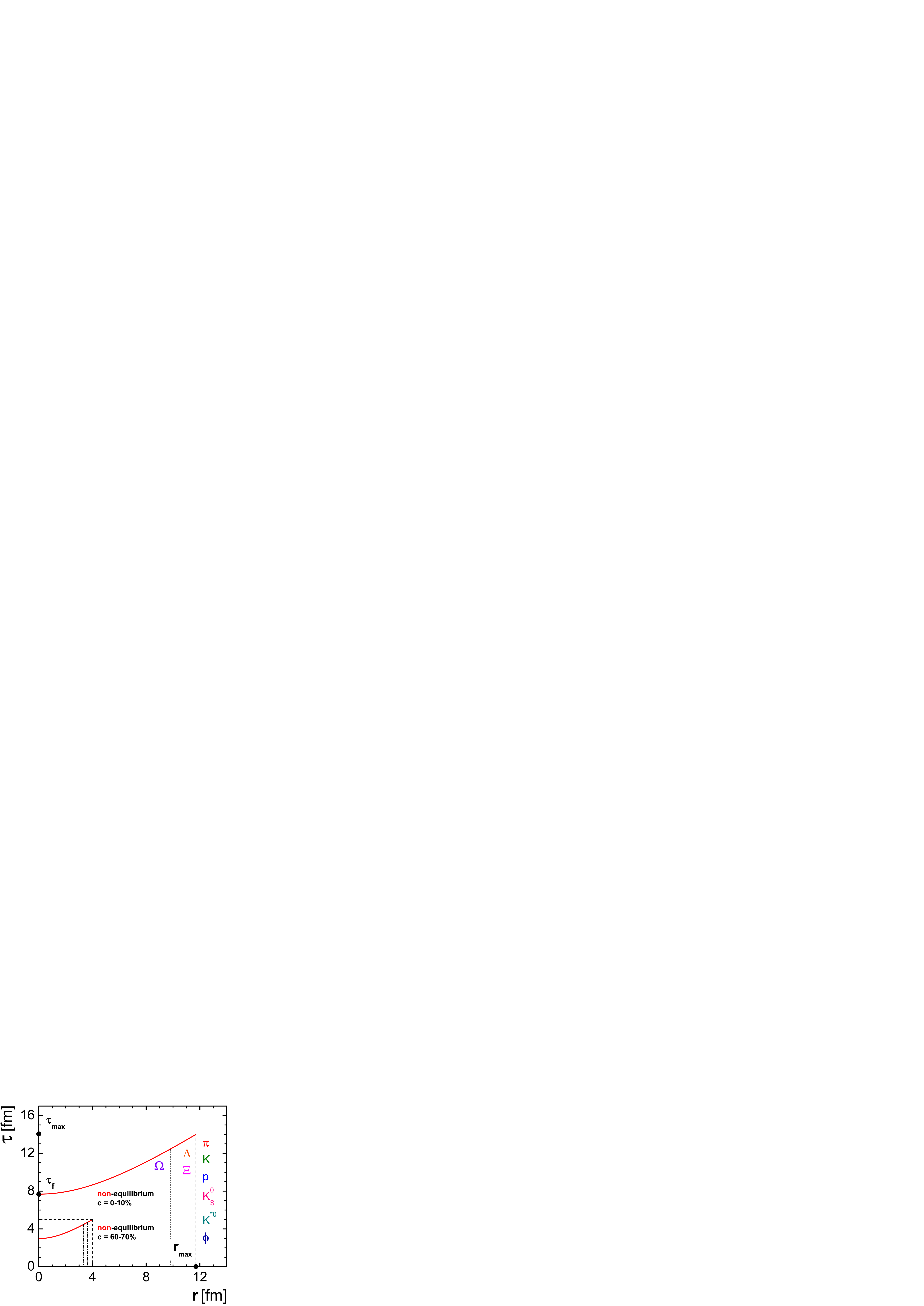}
\end{center}
\caption{(Color online) The freeze-out line in the non-equilibrium
Cracow model.} \label{fig:freezeout}
\end{figure}
%%%%%%%%%%%%%%%%%%%%%%%%%%%%
%
Altogether, our results lead to the freeze-out picture shown in Fig.~\ref{fig:freezeout}. In the center-of-mass frame at $z=0$, the freeze-out starts in the center of the fireball at the time
$\tau_f$. Subsequently, it spreads out along the hyperbola \cite{Kisiel:2005hn,Chojnacki:2011hb}
\begin{eqnarray}
 \tau(r)~=~\sqrt{\tau_f^2+r^2}~\leq~
 \sqrt{\tau_f^2+r_{\rm max}^2}~.
 %\tau^2~=~t^2-z^2~=~\tau_f^2+x^2+y^2~\leq~\tau_f^2+r_{max}^2~.
\end{eqnarray}
The radius $r=\sqrt{0.7} \, r_{\rm max}$ ($r=\sqrt{0.8}\,r_{\rm max}$)
determines the production range for $\Omega$'s and $\Xi$'s ($\Lambda$'s). All other particles are produced in the range ending at $r=r_{\rm max}$.

%%%%%%%%%%%%%%%%%%%%%%%%%%%%%%%%%%%%%%%%%%%%%%%%%%
\section{Conclusions}\label{sect:Discussion}
%%%%%%%%%%%%%%%%%%%%%%%%%%%%%%%%%%%%%%%%%%%%%%

In this work we have analyzed the transverse-momentum spectra of strange hadrons produced in Pb+Pb collisions at the collision energy $\sqrt{s_{\rm NN}}=2.76$~TeV. In this way, we have extended our approach initiated in Ref.~\cite{Begun:2013nga} where we
studied pions, kaons, and protons, only. An additional new aspect of the present work is the complete analysis of the data collected at different centrality classes.

Our approach combines the concept of chemical non-equilibrium with the single-freeze-out scenario. To calculate the transverse-momentum spectra we have used the framework of the Cracow model with thermodynamic parameters established in earlier studies of the ratios of hadron abundances.  The geometric parameters of the model have been obtained from the fit to the pion and kaon spectra.  Using the same thermodynamic and geometric parameters, we have obtained an excellent description of the spectra of $K_S^0$'s, $K^*(892)^0$'s, and $\phi(1020)$'s. These particles have different lifetimes, and the presence of a long hadronic phase after the chemical freeze-out would change the temperature parameters characterizing $K^*(892)^0$'s and $\phi(1020)$'s. Therefore, our simultaneous description of $K^*(892)^0$'s and $\phi(1020)$'s confirms the validity of the single-freeze-out approximation. A satisfactory description is also obtained for $\Lambda$, $\Xi$ and $\Omega$ in the low $p_T$ region. Further improvement of the hyperon spectra may be achieved if we assume that they are
emitted from a smaller, internal part of the system but still at the same thermodynamic conditions.

Our general conclusion is that the chemical non-equilibrium model with essentially  one extra geometric parameter allows for a very good description of the spectra of all measured hadrons in heavy-ion collisions at the LHC energies. Since at lower energies the spectra were very well described by the equilibrium model with $\gamma_q=1$ \cite{Prorok:2006ve}, it may suggest a new physics mechanism of particle production at the LHC.

%%%%%%%%%%%%%%%%%%%%%%%%%%%%%%%%%%%%%%%%%%%%%%%%%%%%%%%%%%%%%%%%%%%%%%%%%%%%%%%%%%%%%%%%%%%%%%%%%%%
\acknowledgments
%%%%%%%%%%%%%%%%%%%%%%%%%%%%%%%%%%%%%%%%%%%%%%%%%%%%%%%%%%%%%%%%%%%%%%%%%%%%%%%%%%%%%%%%%%%%%%%%%%%

V.B and W.F. were supported by Polish National Science Center grant No. DEC-2012/06/A/ST2/00390.  M.R. was supported in part by Polish National Science Center grant No. DEC-2011/01/D/ST2/00772.

\appendix

\begin{table}
{\begin{tabular}{||c||c|c|c|c|c||c|c||}
  \hline
  &  \multicolumn{5}{c||}{ Non-Equilibrium } &  \multicolumn{2}{c||}{ Equilibrium }\\
 \cline{2-8}
    & $r_{\rm max}$   &   $\tau_f$ & $T$ & $\gamma_q$       &  $\gamma_s$     & $r_{\rm max}$    &  $\tau_f$ \\
\hline
~A~ & 12.046          &    7.89          &  137.91          &   1.63          &   2.05           &  11.42           &   9.31 \\
\hline
~B~ & $-1.44\cdot10^{-1}$ &  $-9.34\cdot10^{-2}$ &   $2.57\cdot10^{-5}$ & $-6.17\cdot10^{-5}$ &   $2.05\cdot10^{-3}$ &  $-1.57\cdot10^{-1}$ & $-1.22\cdot10^{-1}$ \\
\hline
~C~ &  $3.97\cdot10^{-4}$ &   $1.65\cdot10^{-4}$ &   $1.08\cdot10^{-6}$ & $-8.34\cdot10^{-6}$ &  $-7.69\cdot10^{-5}$ &   $9.47\cdot10^{-4}$ &  $2.90\cdot10^{-4}$ \\
 \hline
~D~ & $-1.40\cdot10^{-6}$ &   $1.77\cdot10^{-6}$ &  $-6.42\cdot10^{-9}$ &  $5.05\cdot10^{-8}$ &  $-3.56\cdot10^{-7}$ &  $-4.98\cdot10^{-6}$ &  $2.36\cdot10^{-6}$\\
 \hline
\end{tabular}}
\caption{Coefficients used in Eq.~(\ref{ABCD}) describing the centrality dependence of the thermodynamic and geometric parameters for the two versions of the model.} \label{tab:ABC}
\end{table}

%%%%%%%%%%%%%%%%%%%%%%%%%%%%%%%%%%%%%%%%%%%%%%
\section{Rapidity and spacetime rapidity distributions}
\label{app:1}

Using our definition of the element of the freeze-out hypersurface $d\Sigma_\mu$ in (\ref{frye-cooper}) we may write
\begin{eqnarray}
E \frac{dN}{d^3p} = \tau_f \int_0^{r_{\rm max}} r \,dr
\int_{-\infty}^{+\infty} d\eta \int_0^{2\pi} d\varphi \, p \cdot u
f(p \cdot u). \label{ap1e1}
\end{eqnarray}
Integrating this equation over momentum gives
\begin{eqnarray}
\frac{dN}{d\eta} = \tau_f \int_0^{r_{\rm max}} r \,dr
\int_0^{2\pi} d\varphi \, u_\mu \int \frac{d^3p}{E} \, p^\mu f(p
\cdot u). \label{ap1e2}
\end{eqnarray}
The covariant form of the last integral on the right-hand side in (\ref{ap1e2}) implies
\begin{eqnarray}
\frac{dN}{d\eta} = \tau_f \int_0^{r_{\rm max}} r \,dr
\int_0^{2\pi} d\varphi \, n(T,\Upsilon). \label{ap1e3}
\end{eqnarray}
If the freeze-out conditions correspond to constant values of $T$ and $\Upsilon$, the last factor in (\ref{ap1e3}) factorizes and we obtain the desired formula
\begin{eqnarray}
\frac{dN}{d\eta}    = \pi r_{\rm max}^2 \tau_f \, n(T,\Upsilon).
\label{ap1e4}
\end{eqnarray}
In addition, the boost invariance of Eq.~(\ref{ap1e1}) implies that the density $dN/d\eta$ is obtained from the expression whose general form may be written as
\begin{eqnarray}
\frac{dN}{d\eta} = \int dy \int d^2p_T F(p_T,y-\eta),
\label{ap1e5}
\end{eqnarray}
where $F$ is a function of the difference $y-\eta$. From Eq.~(\ref{ap1e5}) we conclude that
\begin{eqnarray}
\frac{dN}{d\eta} = \frac{dN}{dy}.
\label{ap1e6}
\end{eqnarray}

%%%%%%%%%%%%%%%%%%%%%%%%%%%%%%%%%%%%%%%%%%%%%%
\section{Thermodynamic parameters as functions of centrality}
\label{app:2}

In this Section we present our approximate formulas for the centrality dependence of the thermodynamic and geometric parameters in the chemical non-equilibrium and chemical equilibrium models. All functions are approximated by the third-order polynomial of the form
\begin{equation}
A + B c + C c^2 + D c^3,
\label{ABCD}
\end{equation}
where $c$ is given in percentages multiplied by 100. The appropriate coefficients are given in Table \ref{tab:ABC}.

\end{document}